\documentstyle[aps,epsf,preprint]{revtex}
\def \to {\rightarrow}
\def \beq {\begin{equation}}
\def \eeq {\end{equation}}
\def \ba {\begin{eqnarray}}
\def \ea {\end{eqnarray}}
\def \jpsi {J/\psi}
\def \< {\left <}
\def \> {\right >}

\begin{document}
\baselineskip 20pt
\renewcommand{\thesection}{\Roman{section}}
~
\hfill{PKU-TP-97-01}
\vskip 20mm
\begin{center}
{\Large \bf Color-singlet and color-octet $\jpsi$ production in top quark rare decays}
\end{center}
\vskip 10mm

\centerline{Feng Yuan}
\vskip 2mm
\centerline{\small {\it Department of Physics, Peking University, Beijing 100871, People's Republic
of China}} 
\vskip 4mm
\centerline{Cong-Feng Qiao}
\vskip 2mm
\centerline{\small\it China Center of Advanced Science and Technology
(World Laboratory), Beijing 100080, People's Republic of China}
\vskip 4mm
\centerline{Kuang-Ta Chao}
\vskip 2mm
\centerline{\small {\it China Center of Advanced Science and Technology (World Laboratory), Beijing 100080,
People's Republic of China}}
\vskip 1mm
\centerline{\small {\it Department of Physics, Peking University, Beijing 100871, People's Republic of China}} %
\vskip 15mm

\begin{center}
{\bf\large Abstract}

\begin{minipage}{140mm}
\vskip 5mm
\par

$\jpsi$ production in top quark rare decays is investigated under the framework
of NRQCD factorization formalism.
Various production channels are studied, and we find that the contributions
from the color-singlet quark fragmentation and the color-octet gluon
fragmentation are both over $3$ orders larger than that from the leading order
color-singlet process.
The numerical results show that the branching ratio $B(t\to c+\jpsi +X)$
is about $10^{-14}$ in the SM, and $10^{-10}$ in the MSSM.

\vskip 5mm
\noindent
PACS number(s): 14.65.Ha, 14.40.Gx
\end{minipage}
\end{center}
\vfill\eject\pagestyle{plain}\setcounter{page}{1}

Top quark has been discovered by CDF and D0 collaborations at the 
Femilab Tevatron\cite{cdfd0} and its measured mass is around $175 GeV$.
Because it has a mass of the order of the Fermi scale, the top quark may be
strongly associated with the electroweak symmetry breaking (ESB) sector.
So, the studies of top quark (including the production cross section and
the decay widths etc.) may put forward our knowledge about the ESB mechanism.
Furthermore, the fact that both $R_b$ and $A_b$ remain approximately $2\sigma$
away from the SM expectations\cite{rbab} gives us some hints of the probability that
the third generation might couple to some new physics (may be associated with
the ESB mechanism). All these distinguishing features about the top quark and
the third generation require us to measure the top quark properties precisely.

The flavor changing neutral current (FCNC) rare decays of heavy quarks have long
been a subject of intense theoretical and experimental study. The top quark
rare decays have also been investigated in literatures [3-5]. Because in the SM the branching
ratios of the rare decay modes are shown to be very small to be observable, 
($B(t\to c\gamma) \sim 10^{-12}$, $B(t\to c Z) \sim 10^{-12}-10^{-13}$ and
$B(t\to cg) \sim 10^{-10}$\cite{sm}). The FCNC decay modes make an excellent probe for models
beyond the SM. The Two-Higgs-Doublet models give an enhancement to these decay
rates by $3\sim4$ orders of magnitude\cite{sm}. The minimal supersymmetric SM (MSSM) 
also gives the same orders enhancement\cite{susy}.
Recently, some studies of searching for top quark rare decays
at hadron colliders have also been performed\cite{anomalous}.

In searching for rare decay $t\to cg$, a $c$ quark jet and a gluon jet must be
identified, which makes it more difficult than the other two modes
because of the larger background in the hadron collisions.
Otherwise, if the $c$ quark or the gluon hadronizes into heavy quarkonium,
searching for heavy quarkonium in top quark rare decays 
would further identify the top quark effective coupling vertex $tcg$.
For this purpose, in this paper we calculate the $\jpsi$ production rates in top
quark rare decays. 
$\jpsi$ has an extremly clean signature through its leptonic decays into
$e^+e^-$ or $\mu^+\mu^-$, and this makes it very valuable as trigger for
top quark rare decays at high energy hadron colliders.
Furthermore, on the $\jpsi$ production sector, top quark rare decays provide
another facility to study different production mechanisms.
There are $c$ quark and gluon in the decay $t\to cg$, so we can study the $\jpsi$
production via quark fragmentation and gluon fragmentation in top
quark rare decays.

In the past few years, a new factorization formalism of heavy quarkonium
production in high energy collisions has been developed by Bodwin,
Braaten and Lepage in the context of NRQCD\cite{bbl}.
In this approach, the production process is factorized into short and long 
distance parts, while the latter is associated with the nonperturbative
matrix elements of four-fermion operators. 
Addition to the conventional color-singlet mechanism of
heavy quarkonium production, this factorization formalism
provides a new mechanism called color-octet mechanism, in which
the heavy-quark and antiquark pair is produced at short distance in a
color-octet configuration and subsequently
evolves nonperturbatively into physical quarkonium state.
After including this new production
mechanism, one might explain the $\psi^\prime$ ($\jpsi$) surplus
measured by the CDF group at the Tevatron\cite{surplus}.
In the past few years, the applications of the NRQCD factorization formalism
to $\jpsi$($\psi^\prime$) production at various experimental facilities
have been performed\cite{annrev}.

In top quark decays, $\jpsi$ mainly comes from $t\to W^+ b$ followed
by $b$ decays $b\to \jpsi X$ as well as from direct production $t\to bW^+g^*$
followed by $g^*\to \jpsi +X$\cite{direct}.
However, the $\jpsi$ from $b$ decays can be distinguished by using the
so-called Second Vertex Detector, $b$-tagging,
and the directly produced $\jpsi$ can be isolated by studying its
energy spectrum because it has a very soft energy spectrum\cite{direct}.
As shown in the following calculations,
the $\jpsi$ from top quark rare decays mostly come from parton fragmentation
processes in which they have hard energy spectrum.
So, in top quark rare decays, $\jpsi$ is easily detected and distinguished,
and it therefore can be used to observe the top quark rare decay mode $t\to cg$.

According to the NRQCD factorization formalism, the inclusive production of $\jpsi$
in top quark rare decays has the following factorized form,
\beq
\Gamma(t\to c+\jpsi+X)=\sum\limits_n \hat\Gamma(t\to c+(c\bar c)[n]+X)
        \< {\cal O}_n^{\jpsi} \> .
\eeq
Here, $\hat\Gamma_n$ is the short-distance coefficient for producing a $c\bar c$
pair in a configuration denoted by $n$ (including the angular momentum $^{2S+1}
L_J$ and the color index $1$ or $8$). $\< {\cal O}_n^{\jpsi} \> $ is the long-distance
nonperturbative matrix element demonstrating the probability of 
the $c\bar c$ pair in $n$ configuration evolving into a physical charmonium
state $\jpsi$.
$\hat\Gamma_n$ can be calculated perturbatively as an expansion in
coupling constant $\alpha_s$.
Whereas, $\< {\cal O}_n^{\jpsi} \> $ is a
nonperturbative parameter, and practically is determined by fitting to the 
experimental data.
The matrix element $\< {\cal O}_n^{\jpsi} \> $ can be a color-singlet matrix element or
a color-octet matrix element due to the color index number of the $c\bar c$ pair
being $1$ or $8$.
The color-singlet matrix elements are related to the nonrelativistic radial
wave function or its derivatives at the origin, but the color-octet ones
have not the corresponding relations and can only be extracted from experiment now.
For $\jpsi$ production, fortunately the element $\< {\cal O}_8^{\jpsi}({}^3S_1) \> $
has been well determined from large $p_T$ prompt $\jpsi$ production at the Tevatron
\cite{surplus}.

The leading order color-singlet contributions to $\jpsi$ production in top quark
rare decays come from the process displayed in Fig.1(a).
In the Feynman diagrams, the $tcg$ vertex (represented by a black fat dot in Fig.1
and Fig.2) is an effective vertex.
In the SM, this effective vertex can be calculated by loop calculations and the
final result has the following form\cite{sm},
\beq
M_\mu^i=V_{ci}V_{ti}^*[(a_1p^\mu+a_2q_2^\mu+a_3\gamma^\mu)L+
        (b_1p^\mu+b_2q_2^\mu+b_3\gamma^\mu)R],        
\eeq
for an internal quark $i$.
Here $V_{ij}$ is KM matrix element, $L=(1-\gamma_5)/2$, and $R=(1+\gamma_5)/2$.
$q_2$, $q_1$, and $p$ are moment of the top quark, $c$ quark and gluon
associated with the $tcg$ vertex respectively.
The explicit expressions of the six form factors $a_i$ and $b_i$ can be found
in Ref.\cite{sm}.
Adopting this effective $tcg$ vertex in the SM, the decay rates
for the leading order color-singlet process can be calculated straightforward.
The result is lengthy, and it is too tedious to write it here. If we take the limit
$m_c \ll m_t$, we will get 
\ba
\label{e1}
\Gamma^{(LO)}(t\to c+\jpsi)=\frac{2\alpha_s}{243}\frac{\< {\cal O}_1^{\jpsi}({}^3S_1) \> }
        {m_t m_c}(a_3^*a_3+b_3^*b_3).
\ea
In the numerical calculations, we take the input parameters as\cite{sm}
\beq
m_t=175GeV,~~~M_Z=91.2GeV,~~~M_W=80.10 GeV,
\eeq
\beq
\alpha_{em}=1/128.8,~~~\alpha_s(m_t)=0.10,~~~sin^2\theta_w=0.23,
\eeq
\beq
m_u=10MeV,~~~m_s=150MeV,~~~m_c=1,5GeV,~~~m_b=5.0Gev.
\eeq
So, the branching ratio for this process is
\beq
B^{(LO)}(t\to c+\jpsi)=\frac{\Gamma^{(LO)}(t\to c+\jpsi)}{\Gamma(t\to bW^+)}=
        5.4\times 10^{-19},
\eeq
where, the color-singlet matrix element value $\< {\cal O}_1^{\jpsi}({}^3S_1) \> =1.2 GeV^3$
\cite{gf} has been used.
This result is dramatic small, and it is about $8$ orders smaller than
the branching ratio of the inclusive top quark rare decay $t\to cg$
which is about $10^{-10}$\cite{sm}.
This is mainly because the gluon propagator has a suppression factor of order
$O(1/m_t^2)$ due to the large momentum transfer in this process.

Actually, the dominant color-singlet contributions come from the so-called fragmentation
processes shown in Fig.1(b) and Fig.1(c) for quark and gluon fragmentation
respectively. In Fig.1(b), on-shelled $c$ quark (gluon in Fig.1(c)) is
produced from top quark rare decays and then fragmentating into the $\jpsi$.
The contributions from the fragmentation processes mostly come from the region of phase
space in which the $\psi -c$ ($\psi -gg$ in Fig.1(c)) system has
large momentum of order $m_t$ and small invariant mass of order $m_c^2$\cite{gf}\cite{qf}.
In the fragmentation approximation, the decay rates are given by,
for the quark fragmentation\cite{qf},
\ba
\nonumber
\Gamma(t\to \jpsi cg)&=&\Gamma(t\to gc)\times P_{c\to \jpsi}\\
\nonumber
                     &=&\Gamma(t\to gc)\times\frac{16\alpha_s^2(2m_c)}{243}\frac{\< {\cal O}_1^{\jpsi}({}^3S_1) \> }
                        {m_c^3}(\frac{1189}{30}-57{\rm ln} 2)\\
\label{e2}
                     &=&\Gamma(t\to gc)\times 1.2\times 10^{-4} ,
\ea
and for gluon fragmentation\cite{gf},
\ba
\nonumber
\Gamma(t\to \jpsi cgg)&=&\Gamma(t\to cg)\times P^{(1)}_{g\to \jpsi}\\
\nonumber
                      &=&\Gamma(t\to cg)\times  8.28\times 10^{-4}\frac{\alpha_s^3(2m_c)}{m_c^3}
                        \< {\cal O}_1^{\jpsi}({}^3S_1 \> \\
\label{e3}
                        &=&\Gamma(t\to gc)\times 3.2\times 10^{-6} .
\ea
Adding all these contributions Eqs.(\ref{e1}), (\ref{e2}) and (\ref{e3}) together,
we get the total color-singlet contributions to $\jpsi$ production in top
quark rare decays, and the branching ratio is about
\beq
B_{singlet}(t\to c+\jpsi+X)=1.1\times 10^{-14}.
\eeq

The dominant color-octet contributions come from the color-octet gluon
fragmentation process shown in Fig.2. According to the velocity scaling rules,
the color-octet process is suppressed by a factor of order $v^4$, but as shown
in Fig.2, the color-octet process has ${\cal O}(1/\alpha_s^2)$ enhancement
compared with the color-singlet process in Fig.1(c). In the fragmentation
approximation, the decay rates for the color-octet gluon fragmentation process
is given by\cite{gf}
\ba
\nonumber
\Gamma(t\to \jpsi c+X)&=&\Gamma(t\to cg)\times P^{(8)}_{g\to \jpsi}\\
\nonumber
                      &=&\Gamma(t\to cg)\times  1.31\times 10^{-1}\frac{\alpha_s(2m_c)}{m_c^3}
                        \< {\cal O}_8^{\jpsi}({}^3S_1 \> \\
                       &=&\Gamma(t\to gc)\times 1.5\times 10^{-4}  ,
\ea
and the branching ratio is
\beq
B_{octet}(t\to c+\jpsi+X)=1.5\times 10^{-14},
\eeq
where the color-octet matrix element $\< {\cal O}_8^{\jpsi}({}^3S_1 \> =1.5\times 10^{-2}GeV^3$
\cite{gf}.

After including the color-singlet and color-octet contributions, the total
decay rates for $\jpsi$ production in top quark rare decays is
\beq
\label{total}
B_{SM}(t\to c+ \jpsi+X)=2.6\times 10^{-14}.
\eeq
>From the above results, we can see that the $\jpsi$ production in top quark rare
decays is dominated by the color-singlet quark fragmentation and the color-octet
gluon fragmentation, and the contributions from these two processes are both
$4$ orders larger than those from the leading order color-singlet process.
As shown in Ref.\cite{z0}, the energy spectrum of
$\jpsi$ from color-singlet quark fragmentation is hard.
The $\jpsi$s from color-octet gluon fragmentation also
have large energies because the $t\to c+\jpsi$ decay in Fig.2 is a two body
decay process and the energies of the produced $\jpsi$s in this process are
fixed around one half of the top quark mass.
That means the $\jpsi$s produced in top quark rare decays have
much larger energies than those from direct production\cite{direct}.
So, the signals of the $\jpsi$ produced in top quark rare decays are
easily to be detected and distinguished.

In the MSSM, the SUSY QCD correction would provide an enhancement to the top
quark rare decays\cite{susy}. The effective $tcg$ vertex is given in Ref.\cite{susy}, for
an internal squark $\alpha$,
\beq
M_{\mu}^\alpha=-i\frac{\alpha_s}{2\pi}K_{\alpha t}K_{\alpha c}
        (\gamma_\mu P_LV^\alpha+\frac{P_\mu}{m_t}P_RT^\alpha),
\eeq
where $P=q_1+q_2$, and $K_{\alpha q}$ is the supersymmetric version of KM matrix.
The explicit expressions of the form factors $V$ and $T$ can be found in Ref.\cite{susy}.
In the MSSM, the mass eigenstates of squark are obtained by mixing the left- and right-handed
squark with the mixing angle $\theta$. Here, we only consider the unmixing case
($\theta=0$) in which the left-handed squark is a mass eigenstate.
Our calculations can be easily extended to the mixing case.
The mass splitting of the scale top quark and the scale charm quark comes into
account, which is taken to be $m_{\tilde c}=0.9m_{\tilde t}$.
If all scale quark masses would be the same, the decay rates would be identical
to zero.

With this effective $tcg$ vertex, we can calculate the $\jpsi$ production rates
in top quark rare decays by using the same way discussed above.
The decay rates for the leading order color-singlet process Fig.1(a) is
\ba
\nonumber
\Gamma_{SUSY}^{(LO)}(t\to c+\jpsi)&=&
        \frac{\alpha_s^3 \epsilon^2}{486\pi^2} \frac{\< {\cal O}_1^{\jpsi}({}^3S_1 \>  }
        {(m_t^2-m_c^2)^2 m_c m_t^3}[ - 36m_c^6T^2 + 24m_c^4m_t^2T^2 - 24m_c^4m_t^2TV \\
        &~&- 27m_c^4m_t^2V^2 + 12m_c^2m_t^4T^2 + 24m_c^2m_t^4TV + 2m_c
        ^2m_t^4V^2 + m_t^6V^2],
\ea
where $V=V^{\tilde t}-V^{\tilde c}$ and $T=T^{\tilde t}-T^{\tilde c}$.
Here $\epsilon$ is a small number appearing in the SUSY-KM matrix, which is
taken as $\epsilon^2={1\over 4}$\cite{susy}.
For the quark and gluon fragmentation (including color-singlet and color-octet
contributions), the branching ratio is
\ba
B_{SUSY}^{(frag)}(t\to c+\jpsi+X)=B_{SUSY}(t\to cg)[P_{c\to \jpsi}+
        P_{g\to \jpsi}^{(1)}+P_{g\to \jpsi}^{(8)}]
\ea

All the above results are summarized in Fig.3 and Fig.4.
In Fig.3, we demonstrate the dependence of $\jpsi$ production branching
ratio on the scale top quark mass $m_{\tilde t}$ in the case that
$m_{\tilde g}=150GeV$. The dotted line represents the contribution from
leading order color-singlet process, the dashed line from color-singlet
quark and gluon fragmentation processes, and the dotted-dashed line from
color-octet gluon fragmentation process.
In Fig.4, we plot the branching ratio as a function of the gluino
mass $m_{\tilde g}$ in the case that $m_{\tilde t}=200GeV$.
The curves in Fig.4 represent the same things as those in Fig.3.
>From these two figures, we can see that similar to that in the SM the $\jpsi$
production is dominated by the partons fragmentations (including the color-singlet
and color-octet contributions) in the MSSM.
Except for some lower stop mass region ($<150GeV$) the branching ratio of
$B(t\to c+\jpsi+X)$ is not sensitive to the stop quark mass and the gluino mass.
Compared with the results obtained in the SM Eqs.(\ref{total}), it is shown
that the SUSY QCD contributions can enhance the $\jpsi$ production rates in
top quark rare decays by over $4$ orders, and the branching ratio of
$B(t\to c+\jpsi+X)$ is now up to about $10^{-10}$.
But it is still too small to be observable.

In conclusion, in this paper we have calculated $\jpsi$ production rates in top
quark rare decays. We find that both in the SM and in the MSSM the contributions
from partons fragmentations (color-singlet quark fragmentation and color-octet
gluon fragmentation) are over $3$ orders larger than that from leading order
color-singlet process.
However, the total production rates of $\jpsi$ in top quark rare decays are
too small to observable, {\it i.e.}, $10^{-14}$ in the SM and $10^{-10}$ in the
MSSM.
So, the new physics effects would be crucial important to this decay mode,
such as some anomalous top quark interactions.
We hope that the searching for $\jpsi$'s signals in top quark rare decays will
be hold at the top quark factories such as the {\bf LHC} and {\bf NLC}.

\vskip 1cm
\begin{center}
\bf\large\bf{Acknowledgements}
\end{center}

This work was supported in part by the National Natural Science Foundation
of China, the State Education Commission of China and the State Commission
of Science and Technology of China.


\newpage
\centerline{\bf \large Figure Captions}
\vskip 2cm
\noindent
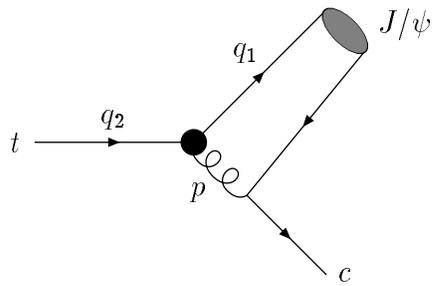
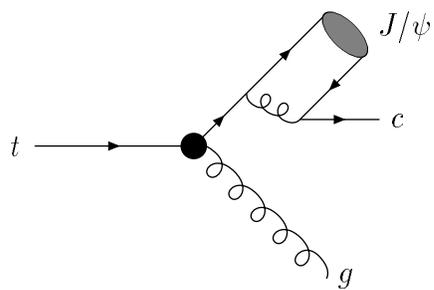
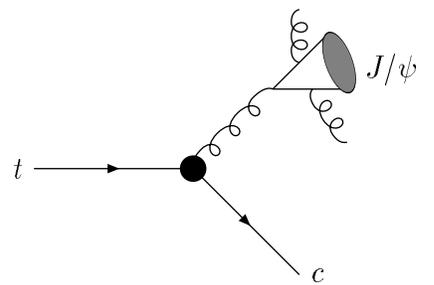
Fig.1. Color-singlet processes in $\jpsi$ production in top quark rare decays.
(a) The leading order process; (b) quark fragmentation process; (c) gluon
fragmentation process.

\noindent
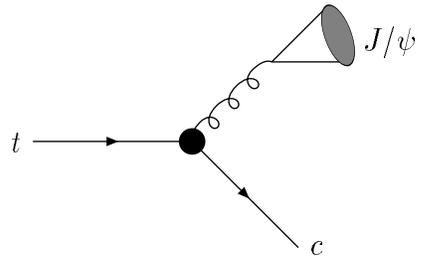
Fig.2. Color-octet gluon fragmentation process in $\jpsi$ production in top
quark rare decays.

\noindent
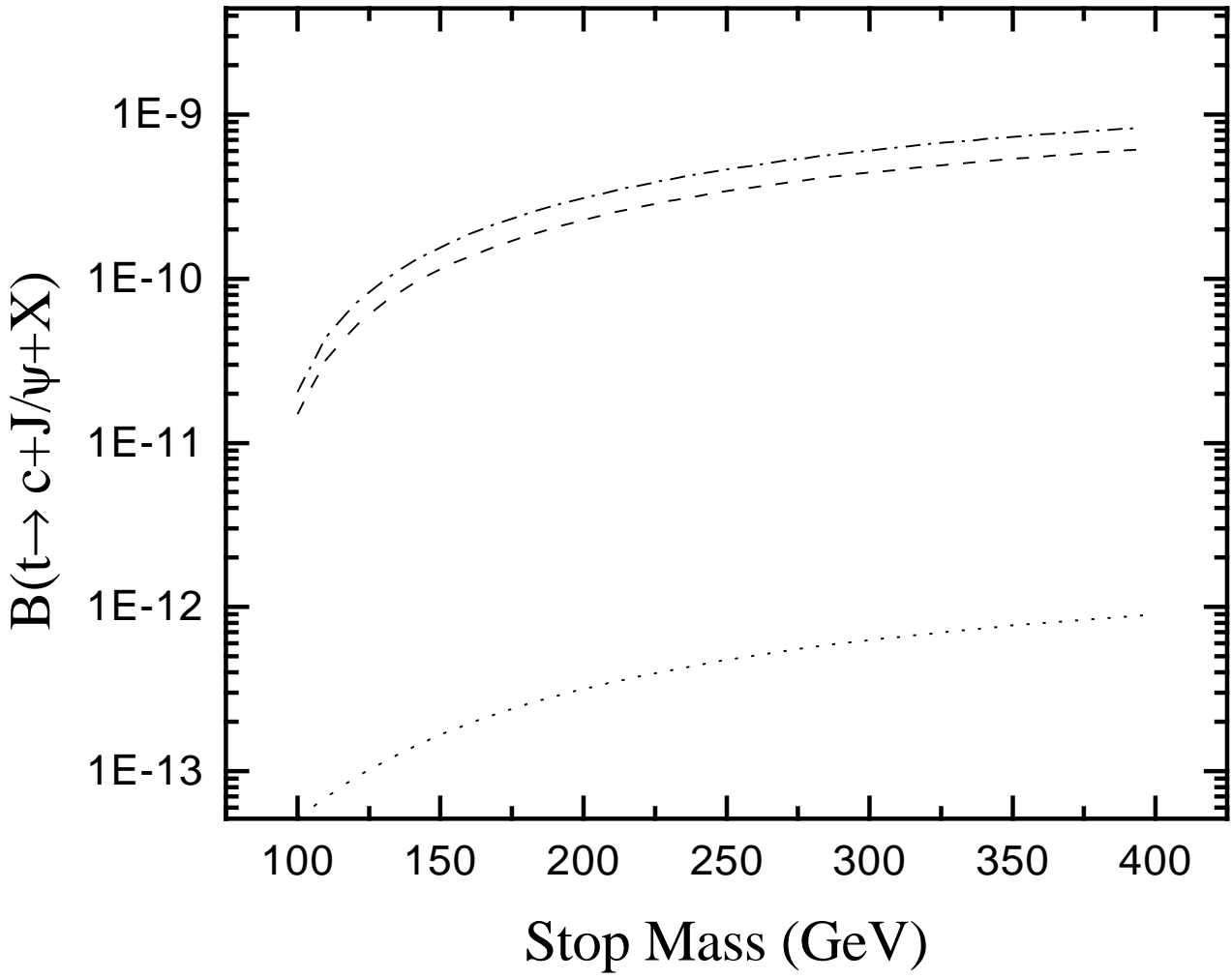
Fig.3. The branching ratio $B(t\to c+\jpsi+X)$ as a function of the scale top quark mass
in the MSSM. The dotted line represents the contribution from the leading order
color-singlet process, the dashed line from the color-singlet quark and gluon
fragmentation processes, and the dotted-dashed line from the color-octet gluon
fragmentation process.

\noindent
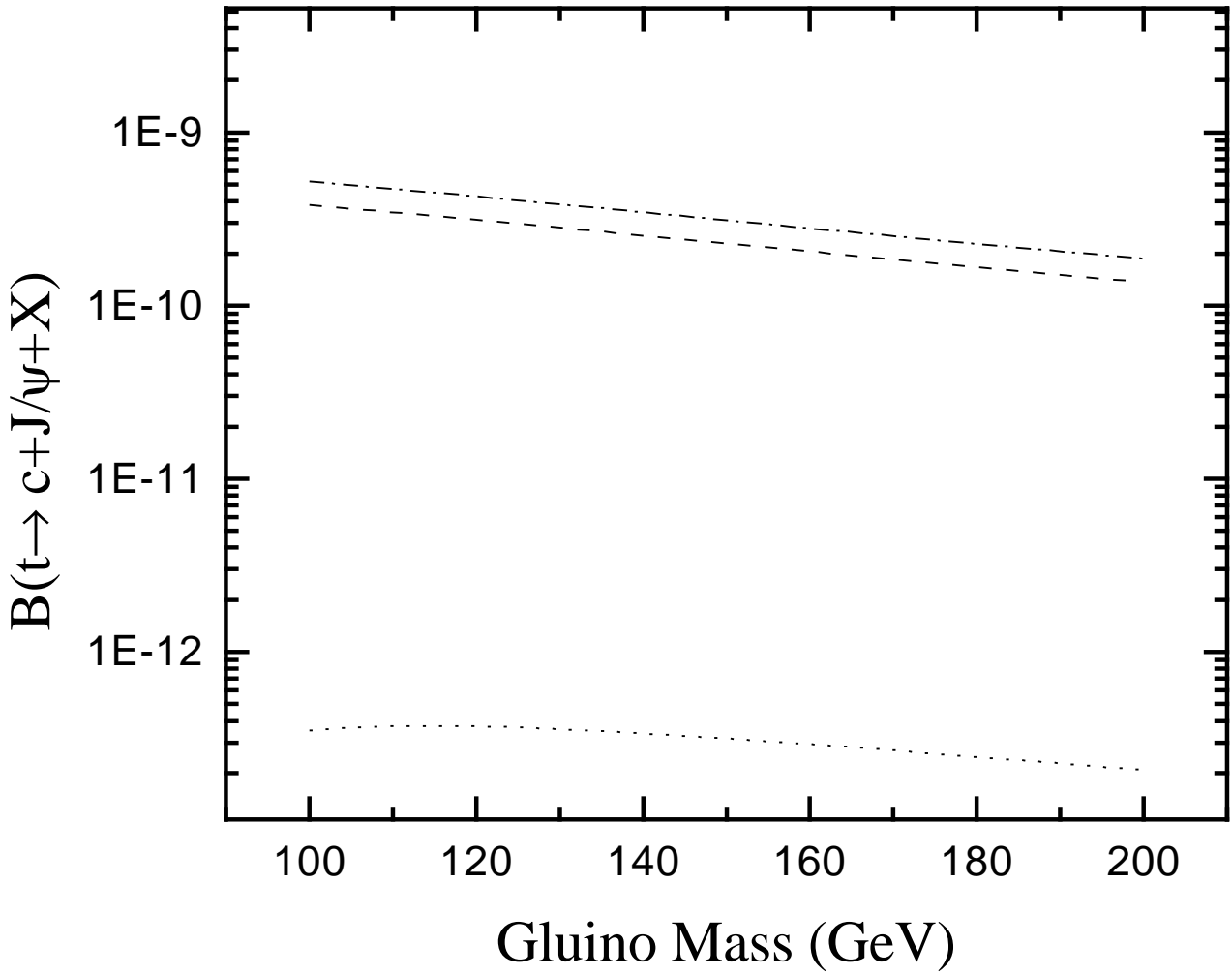
Fig.4. The branching ratio $B(t\to c+\jpsi+X)$ as a function of the gluino
mass in the MSSM. The meanings of the curves are as the same as those
in Fig.3.

\end{document}